\documentclass[reprint,aps,prb]{revtex4-2}
\usepackage{graphicx}
\usepackage{amsmath}
\usepackage{amssymb}
\usepackage{dcolumn}
\usepackage{bm}
\newcommand{\bfm}[1]{\textbf{\em#1}}

\begin{document}

\preprint{APS/123-QED}

\title{Simulation of bright and dark diffuse multiple scattering lines in high-flux synchrotron X-ray experiments}

\author{Maur{\'{i}}cio B. Estradiote}
\affiliation{Institute of Physics, University of S{\~{a}}o Paulo, S{\~{a}}o Paulo, SP, Brazil}

\author{A. G. A. Nisbet}
\email{Gareth.Nisbet@diamond.ac.uk}
\affiliation{Diamond Light Source, Harwell Science \& Innovation Campus, Harwell OX11 0DE, United Kingdom}

\author{Rafaela F. S. Penacchio}
\affiliation{Institute of Physics, University of S{\~{a}}o Paulo, S{\~{a}}o Paulo, SP, Brazil}

\author{Marcus A. R. Miranda}
\affiliation{Department of Physics, Federal University of Cear{\'{a}}, Fortaleza, CE, Brazil}

\author{Guilherme A. Calligaris}
\affiliation{Brazilian Synchrotron Light Laboratory - SIRIUS/CNPEM, Campinas, SP, Brazil}

\author{S{\'{e}}rgio L. Morelh{\~{a}}o}
\email{morelhao@if.usp.br}
\affiliation{Institute of Physics, University of S{\~{a}}o Paulo, S{\~{a}}o Paulo, SP, Brazil}

\begin{abstract}
We present a theoretical framework for understanding diffuse multiple scattering (DMS) in single crystals, focusing on diffuse scattering-Bragg (DS-Bragg) channels. These channels, when probed with high-flux, low-divergent monochromatic synchrotron X-rays, provide well-defined visualizations of Kossel lines. Our main contribution lies in modeling the intensity distribution along these lines by considering DS around individual reciprocal lattice nodes.  The model incorporates contributions from both general DS and mosaicity, elucidating their connection to second-order scattering events. This comprehensive approach advances our understanding of DMS phenomena, enabling their use as probes for complex material behavior, particularly under extreme conditions.
\end{abstract}

\maketitle

\section{Introduction}
X-ray diffuse scattering (DS) in crystals arises from any deviation from perfect periodicity, encompassing phenomena like atomic thermal vibrations, defects, and local disorder \cite{pd13,iw28,yw31,wz40,bw90}. Thermal DS contains information about lattice dynamics of the materials and, historically, it was the first tool utilized for experimental determination of phonon dispersion relations \cite{po48,hc52,rj54,ej55}. The advent of synchrotron radiation technology has not only revitalized DS as a feasible probe for studying phonons \cite{mh99,rx05,am15}, but has also transformed it into a powerful technique for investigating a wide range of order/disorder related phenomena \cite{tw10,ri12,mk12,nr21,kh21,es22,kt22,nw23,ro23,xg23,tb23,ds23,mz23}. Synchrotron X-rays of very high flux, approaching $10^{16}\,{\rm ph}/{\rm s}/{\rm mm}^2$, have also revealed relatively unknown second-order scattering processes between DS sources and Bragg reflections. While the Bragg-DS channel produce nebulous and poorly localized intensity distributions, the DS-Bragg channel, where X-rays originating from DS undergo subsequent Bragg diffraction, yields well-defined intensity lines \cite{ir09,an15}. The diffuse multiple scattering (DMS) lines, as they have been called, appear as a kind of Kossel lines \cite{rt70,sm91}, but without using divergent beam X-ray generators or any instrumental artefacts to produce divergent beams of monochromatic X-rays. Conversely, DMS lines become visible only with very narrow and highly parallel monochromatic beams, arising primarily from DS sources within the crystal. 

Increasingly accessible with advanced X-ray sources and detectors, DMS lines are more than a nuisance in diffraction experiments. They offer unique insights into crystal defects, strain fields, and lattice dynamics, complementing traditional diffraction techniques \cite{an21,an23}. The growing interest in utilizing DMS lines stems from their sensitivity to subtle crystallographic changes, coupled with the ability to monitor these changes across many directions simultaneously by analyzing scattered X-rays within a single, relatively small solid angle. This capability makes DMS lines a powerful probe in experiments with limited instrumental degrees of freedom, such as those conducted under extreme conditions. To fully exploit this potential, a comprehensive approach for structure modeling based on DMS lines is essential. Determining the geometric positions of these lines from projection of crystal's Bragg Cones (BCs) is relatively straightforward with existing software tools, like the \texttt{klines} module in the PyDDT package \cite{rp23}. The true difficulty lies in accurately describing the intensity distribution along the lines. This work introduces a theoretical framework to tackle the DMS line intensity analysis challenge.

\section{Theoretical Framework}\label{ss:theoaspect}

\begin{figure}
\includegraphics[width=.35\textwidth]{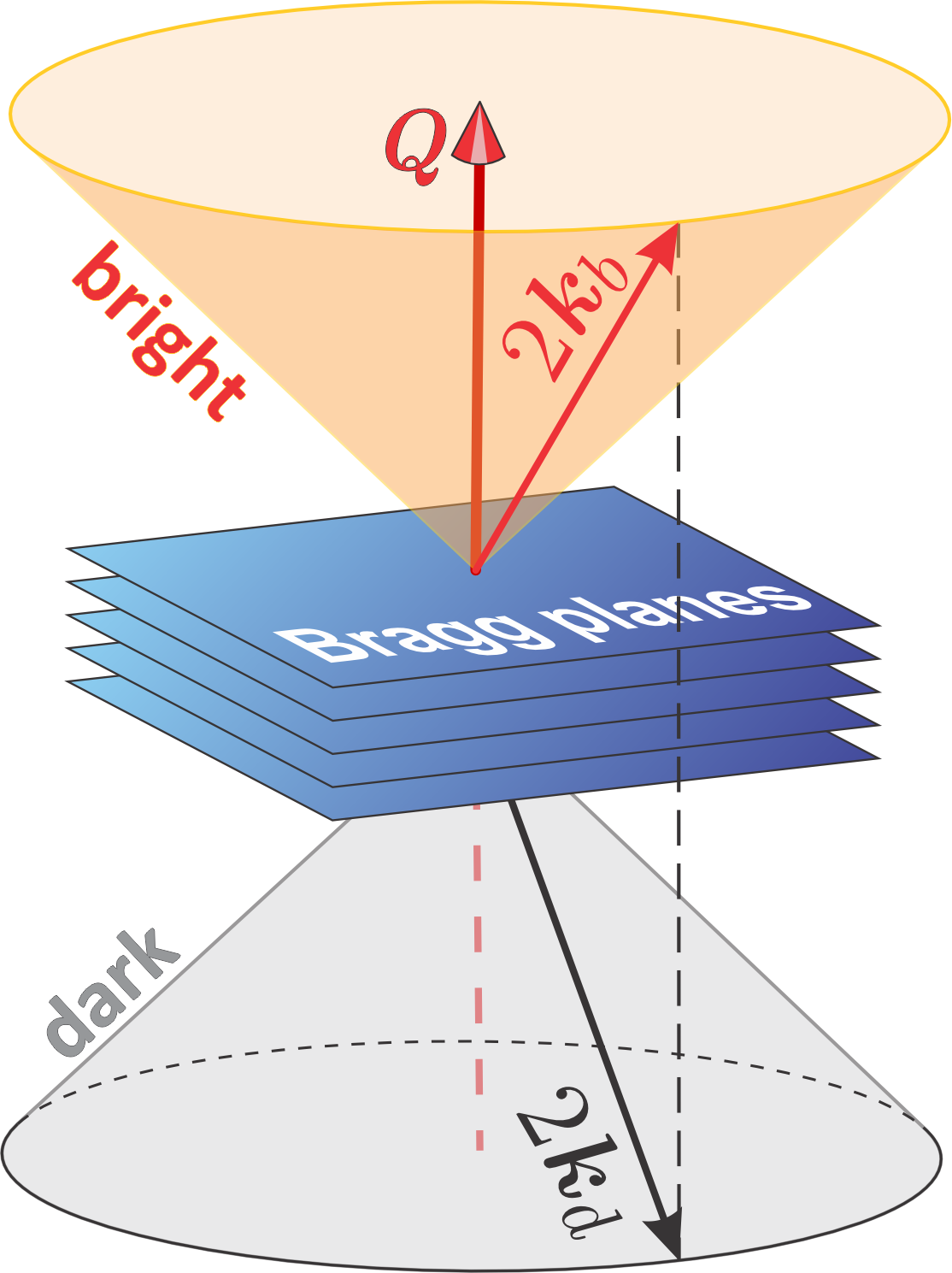}\\
\caption{Bright and dark BCs for diffraction vector ${\bfm Q}$. X-rays with wavevector $\bfm{k}_{d}$ propagating along the dark cones are attenuated by diffraction as they are scattered over the bright cone with wavevector $\bfm{k}_{b} = \bfm{Q} + \bfm{k}_{d}$. }
\label{fig:bcones} 
\end{figure}

One approach for accomplishing this task is to further exploit the concept of BCs, but distinguishing between two sets of wavevectors that are related to a reciprocal lattice vector $\bfm{Q}$. In Fig.~\ref{fig:bcones} these two sets are depicted, bright BCs as representing the set of wavevectors $\bfm{k}_{b}$ that fulfill the condition $2\bfm{k}_{b}\cdot\bfm{Q}= \left \vert \bfm{Q} \right \vert ^2$, while dark BCs represent the wavevectors $\bfm{k}_{d}$ that fulfill $2\bfm{k}_{d}\cdot\bfm{Q}= -\left \vert \bfm{Q} \right \vert ^2$ where $\bfm{k}_{b} = \bfm{Q} + \bfm{k}_{d}$. 

\begin{figure}
\includegraphics[width=.48\textwidth]{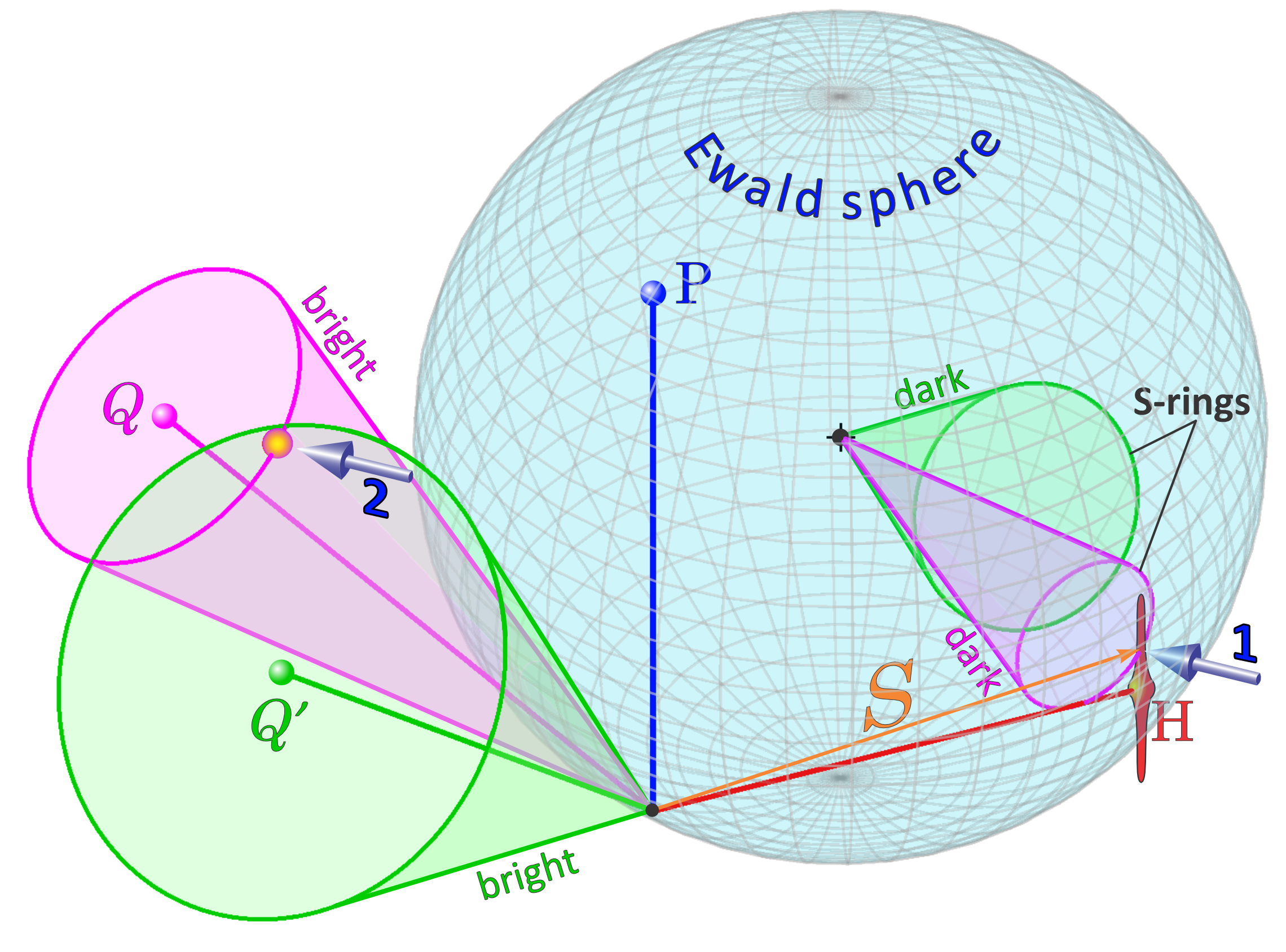}\\
\caption{Ewald construction of reciprocal space where dark cones have their origin at the center of the Ewald sphere. Every scattering vector $\bfm{S}$ touching the cone-sphere intersection (S-ring) gives intensity to a corresponding spot over the bright cone centered at the origin of reciprocal space. In this illustration, the S-ring of reflection Q touches (arrow 1) an intensity distribution near the reciprocal node of reflection H, defining the $\bfm{S}$ vector of the main intensity contribution (arrow 2) over the Q bright cone. It occurs close to the bright cone of reflection Q$^\prime$, although the S-ring of reflection Q$^\prime$ is quite far from the H node.}
\label{fig:ewaldconstruction} 
\end{figure}

In the Ewald construction of reciprocal space, as illustrated in Fig.~\ref{fig:ewaldconstruction}, each dark cone originates at the center of the Ewald sphere, implying that the dark cones move as the direction of the incident wavevector $\bfm{k}$ changes. The intersection of the dark cone with the surface of the Ewald sphere defines a ring containing all possible elastic scattering vectors $\bfm{S} = \bfm{k}_{d} - \bfm{k}$, capable of providing intensity contribution to the bright cone through diffraction vector $\bfm{Q}$. Bright cones are fixed at the origin of reciprocal space, as their positions are independent of the incident beam direction. The intensity along a DMS line---the projection of the bright cone onto the detector area---is influenced by three key factors: \textit{i)} Proximity of the S-ring (the set of scattering vectors $\bfm{S}$) to reciprocal lattice nodes: closer proximity to a node results in higher intensity at specific points along the line. \textit{ii)} Intensity distribution around each node: the unique intensity pattern surrounding each node creates peculiar features in the line. \textit{iii)} Amount of DS between nodes: DS contributes to the overall visibility of the DMS line. Fig.~\ref{fig:ewaldconstruction} shows an example where the dark cone of reflection Q intercepts the Ewald sphere close to the node of reflection H. As the corresponding S-ring probes the node's nearby intensity, an enhancement of intensity can be seen over the DMS line of reflection Q. When the S-ring touches the node, there is a vector $\bfm{S}$ coinciding with the node's reciprocal lattice vector $\bfm{H}$ and, in this case, multiple diffraction (MD) dynamical theory \cite{sc84,ew97,aa08} is necessary to properly describe the intensities of the strongly coupled X-ray waves propagating inside the crystal; one wave from the reflection H and another from the reflection ${\rm P} = {\rm H} + {\rm Q}$. 

Aiming to treat DMS intensities that are visible away from strong Bragg reflections, the second-order solution of the MD \cite{qs00,sm02a,qs06} provides a suitable approach for describing situations where the wavefield $\bfm{D}_1$ of the P reflection, usually referred to as the primary reflection, is much weaker than the wavefield $\bfm{D}_2$ from the sequence of H and Q reflections, also know as the \textit{Umweg} wave \cite{ys00,sm01}. When the $\bfm{S}$ vectors are apart from any reciprocal lattice node, the wavefield inside the crystal, $\bfm{D} = \bfm{D}_1 + \bfm{D}_2$, simplifies to 
\begin{eqnarray}\label{eq:detourwave}
    \bfm{D} \simeq \bfm{D}_2 & = & R(\bfm{H}) F_{\rm H} F_{\rm Q} \left(\hat{\bfm{k}}_b \! \times \! \left\{ \hat{\bfm{k}}_b \! \times \! \left[\hat{\bfm{k}}_d \! \times \! \left( \hat{\bfm{k}}_d \! \times \! \bfm{D}_0 \right) \right] \right\} \right) \nonumber \\
     & = & D_0 R(\bfm{H}) F_{\rm H} F_{\rm Q}\, \boldsymbol{\varepsilon}\,.
\end{eqnarray}  
In standard MD treatments concerning with phase measurements by the interference between the $\bfm{D}_1$ and $\bfm{D}_2$ wavefields \cite{qs06,sm11,az14,sm15,sm17,av20,rp23}, the resonance term $R(\bfm{H})$ determines the excitement of the \textit{Umweg} wave as a function of the H node distance to the surface of the Ewald sphere or, equivalently, to the S-ring. $F_{\rm X}$ stands for the structure factor of reflection X (= H or Q), and the directions $\hat{\bfm{k}}_d$ and $\hat{\bfm{k}}_b$ of the wavevectors on the dark and bright cones of the Q reflection provide the polarization factor $\boldsymbol{\varepsilon}\cdot\boldsymbol{\varepsilon}^*$, computed with respect to the state of polarization of the incident wave $\bfm{D}_0 = D_0 \hat{\boldsymbol{\varepsilon}}_0$. 

An expression for simulating the intensity distribution along the DMS line of a given reflection Q follows directly from Eq.~(\ref{eq:detourwave}) as
\begin{equation}\label{eq:DMSlineintensity}
    I_{\rm Q}(\bfm{S}) = I_0 \left| F_{\rm Q} \right|^2 \boldsymbol{\varepsilon}\cdot\boldsymbol{\varepsilon}^* \sum_{\rm H}{ \left| F_{\rm H} \right|^2 W(\bfm{S}-\bfm{H})}
\end{equation}
where the previous resonance term is replaced by a more general function $W(\bfm{S}-\bfm{H})$ to take into account elastic scattering that occurs away from exact Bragg conditions. Within this simple approach, the same function $W$ applies to all H nodes, and the DS intensities are describable as a discrete superposition of contributions from each individual H node. However, this approach can be easily extended to more complex situations that require a particular $W$ function per node to proper describe DS in the whole reciprocal space accessible by a given S-ring.

To compute the relative widths of observable DMS lines, the dynamical intrinsic width $\Lambda_\theta$ of the reflectivity curve in specular scattering geometry is used as reference \cite{aa08,an11,sm16}. In reciprocal space, it is accounted for as a variation in the modulus $Q$ of the diffraction vector, that is, $Q^2 \rightarrow (Q \pm \Lambda/2)^2 \simeq Q^2 \pm Q\Lambda$ where $\Lambda = (4\pi/\lambda) \cos\theta_B \Lambda_\theta$ and $\theta_B$ is the Bragg angle of reflection Q. It results in a selection criterion for the full set of wavevectors on the bright BC, in which
\begin{equation}\label{eq:bcwidth}
    \left|2\bfm{k}_b \cdot \bfm{Q} - Q^2 \right| \leq \frac{1}{2}Q\Lambda = \frac{8 \pi^2 r_e}{3 V_{\rm cell}} \left|F_{\rm Q}F_{\bar {\rm Q}}\right|^{1/2}
\end{equation}
where $r_e = 2.818\times10^{-5}$\,\, is the classical electron radius and $V_{\rm cell}$ is the unit cell volume. In practice, Eq.~(\ref{eq:bcwidth}) implies that the width of DMS lines are proportional to the structure factor modulus of the corresponding Q reflection as $F_{\bar{\rm Q}}$ refers to the -Q reflection and $|F_{\rm Q}F_{\bar {\rm Q}}|^{1/2}\simeq|F_{\rm Q}|$. Note that the full set of wavevectors on the bright BC of reflection Q match, exactly, the set of wavevectors on the dark BC of reflection -Q. Consequently, when DS intensities from primary sources are directly measurable, dark BCs can become visible as shadows against such diffuse signals. In other words, the intensity along a DMS line can appear, in principle, either as positive or negative with respect to the existing scattered intensity in the detector area.

Intensity simulation of DMS lines implies in modeling the three-dimensional intensity distribution around each node. Besides general DS sources such as point defects, stacking faults, atomic thermal vibrations, and many other types of deviation from the average periodic structure \cite{mh99,rx05,am15,nr21,nw23,ro23}, crystal truncation is a deviation from infinite periodicity and must be taken into account as extended scattering sources close to the nodes in addition to DS sources. As a first approximation to identify experimental condition favorable to the visibility of DMS lines, isotropic DS centered on each node and truncation rods in crystal slabs of outwards surface normal direction $\hat{\bfm{z}}$ are modeled as follow. 
\begin{equation}\label{eq:W}
    W(\bfm{u}) = \frac{\eta_1}{1 + (\pi N u)^2} + \frac{\eta_2}{1+(\pi N_{xy} u_{xy})^2} \left | \frac{\sin(\pi N_z u_z)}{\pi N_z u_z}\right|^2 
\end{equation}
in Eq.~(\ref{eq:DMSlineintensity}) is written in terms of the distance 
\begin{equation}\label{eq.u}
    \bfm{u} =\left[\left ({\rm h}^\prime-{\rm h}\right)\bfm{a}^* + \left ({\rm k}^\prime-{\rm k}\right)\bfm{b}^* + \left ({\rm l}^\prime-{\rm l}\right)\bfm{c}^*\right]/\langle L^* \rangle
\end{equation}
from each node, that is, from each reflection H of indeces hkl. The non integer indeces of the scattering vectors $\bfm{S}$ are obtained as ${\rm h}^\prime=\bfm{S}\cdot\bfm{a}/2\pi$, ${\rm k}^\prime=\bfm{S}\cdot\bfm{b}/2\pi$, and ${\rm l}^\prime=\bfm{S}\cdot\bfm{c}/2\pi$ through the unit cell edge vectors $\bfm{a}$, $\bfm{b}$, and $\bfm{c}$, while $\langle L^* \rangle^3 = \bfm{a}^*\cdot(\bfm{b}^*\times\bfm{c}^*)$ is the reciprocal unit cell volume. The extent $\langle L^* \rangle/(\pi N)$ of isotropic DS from the nodes is adjustable by the number $N$. $\eta_1$ and $\eta_2$ weight the DS and crystal truncation rod (CTR) contributions in the case of a thin slab of thickness $N_z$, in number of unit cells of mean edge $2\pi/\langle L^* \rangle$. For slabs thicker than the beam coherence length and/or X-ray penetration depth, the sinc-squared function is replaceable by its enveloping Lorentzian function $1/[1 + (\pi N_z u_z )^2]$ and $N_z$ becomes an effective number, as the one in the other term regarding the effective in-plane slab dimension $N_{xy}$ with $u_{xy} = \left| \bfm{u}-\bfm{u}_z \right|$ and $\bfm{u}_z = (\bfm{u}\cdot\hat{\bfm{z}})\hat{\bfm{z}}$. For nodes that are anisotropic regarding in-plane $\hat{\bfm{x}}$ and $\hat{\bfm{y}}$ orthogonal directions, this last term gives place to other functions having, for instance, $\pi N_x u_{x}$ and $\pi N_y u_y$ as arguments where $u_x = \bfm{u}\cdot\hat{\bfm{x}}$ and $u_y = \bfm{u}\cdot\hat{\bfm{y}}$. 

\section{Results and Discussions}

\begin{figure}
\includegraphics[width=.48\textwidth]{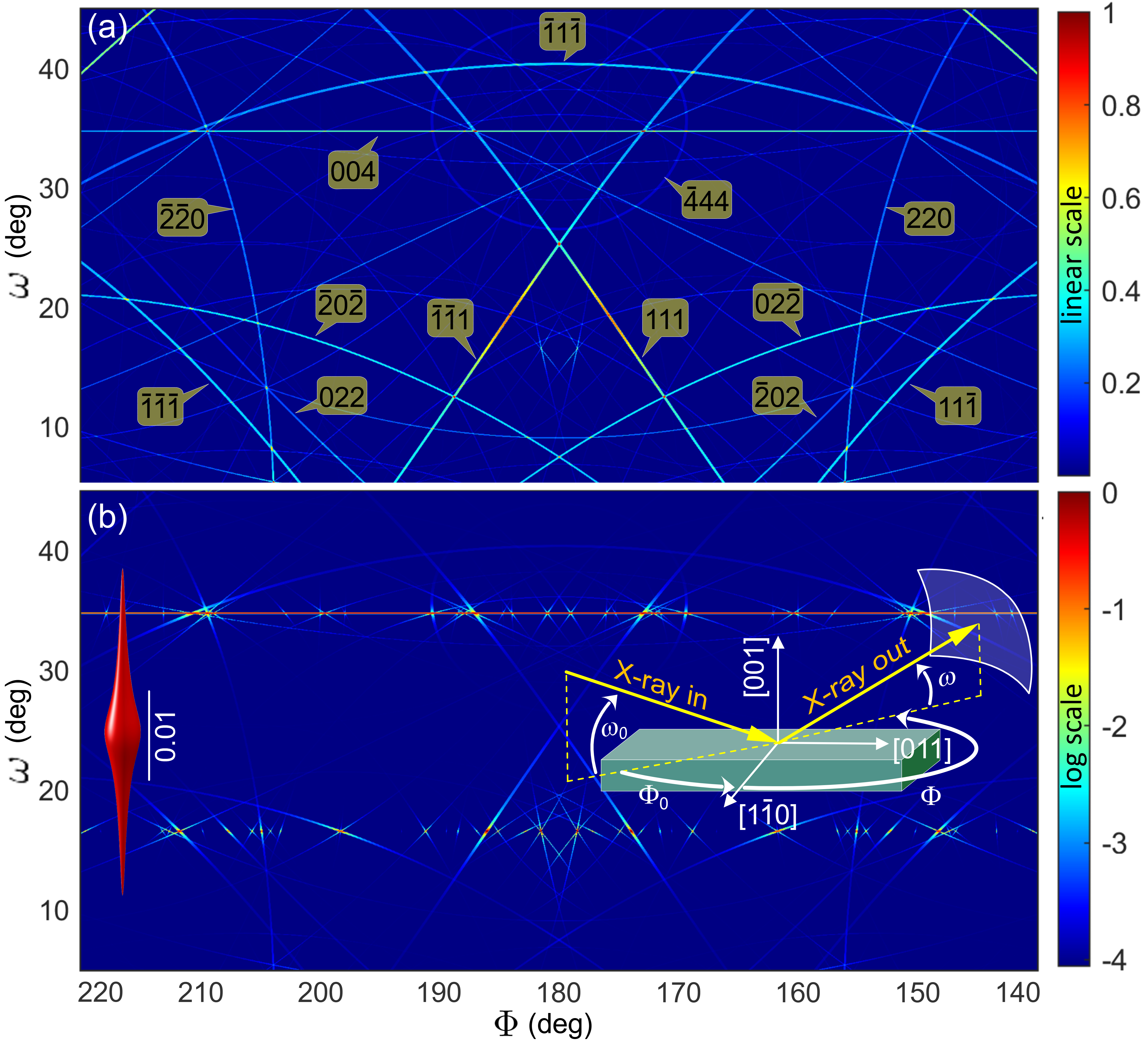}\\
\caption{Simulation of bright DMS lines in silicon (001) for 8\,keV $\sigma-$polarized X-rays. (a) Isotropic DS only, $\eta_1=1$, $\eta_2=0$, and $N=1000$ in $W(\bfm{u})$, Eq.~(\ref{eq:W}). Line indexing by \texttt{klines} \cite{rp23}. (b) Anisotropic intensity distribution around the nodes, elongated along the CTR (left inset, isosurfarce at 3\% of the maximum, scale bar along the L reflection index), no isotropic DS contribution: $\eta_1=0$, $\eta_2=1$, $N_{xy} = 1000$ and $N_z = 100$ in $W_(\bfm{u})$. Incident X-ray direction $\omega_0 = 16.58^\circ$ (002 Si forbidden reflection) and $\Phi_0 = 0$, away from MD cases in the chosen reference frame (right inset). In both images (top and bottom panels), the $\Phi$-axis sense has been reversed to match the sample's perspective. Image resolution of $0.04^\circ$ (pixel size).}
\label{fig:simSilicon} 
\end{figure}

\begin{figure}
\includegraphics[width=.48\textwidth]{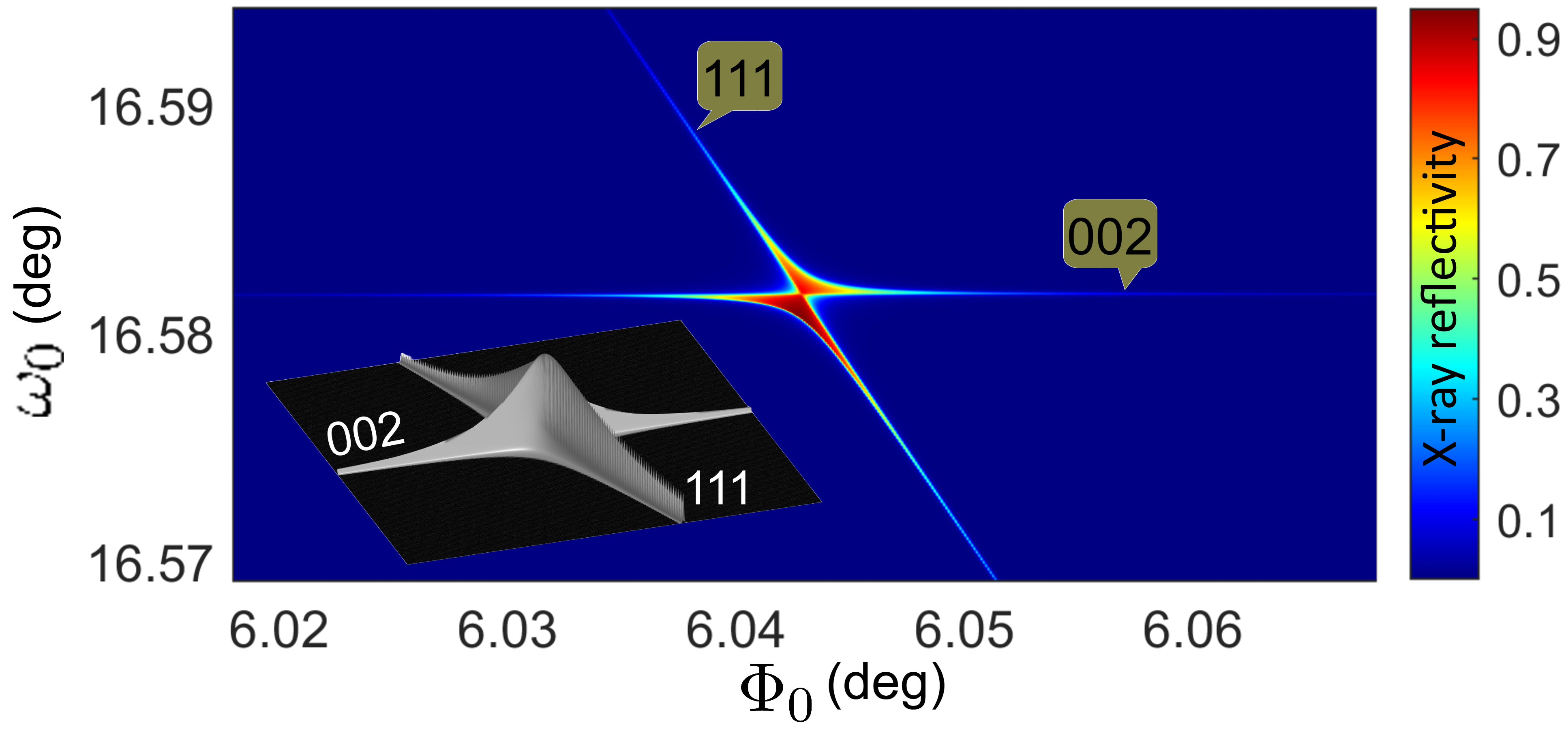}\\
\caption{X-ray reflectivity of the 002 Si forbidden reflection as a function of the $\omega_0$ and $\Phi_0$ angles of incidence [right inset in Fig.~\ref{fig:simSilicon}(b)] around MD with secondary 111. Dynamical diffraction theory for X-rays of 8\,keV in 100\,$\mu$m thick Si (001) slab. Inset: line profiles along the 002 and 111 BCs in linear scale. Experimental data are reported elsewhere \cite{sm02b,jd16}. }
\label{fig:meshscan} 
\end{figure}

Fig.~\ref{fig:simSilicon} shows simulated DMS lines in silicon where the line widths are proportional to $|F_{\rm Q}|$, Eq.~(\ref{eq:bcwidth}), but adjustable by a proportionality constant to make them visible within the resolution of the wide solid angle considered in the simulations. An isotropic DS with lorentzian-like profile contributes to the visibility of all lines with very smooth intensity variation along each line, as seen in Fig.~\ref{fig:simSilicon}(a). In the absence of multiple diffraction, the lines exhibit high contrast on a linear scale regardless of X-ray incidence angle. On the other hand, the anisotropic nature of the CTR severely hampers line visibility, requiring a logarithmic intensity scale for visualization, as in Fig.~\ref{fig:simSilicon}(b) where enhanced intensity contributions are well localized on specific points along the lines and the position of such points very susceptible to the incidence direction. In both cases, Figs.~\ref{fig:simSilicon}(a) and \ref{fig:simSilicon}(b), no Bragg reflection is directly excited by the incident X-rays, that is, the primary wavefield $\bfm{D}_1$ is null. The incidence angle $\omega_0$ was set to the Bragg angle of 002 Si forbidden reflection, whose reflectivity is observable only very near MD cases, as shown in Fig.~\ref{fig:meshscan} by MD dynamical theory calculation for a thick crystal slab \cite{rc74,qs93}. Although, intersection of DMS lines point out incidence directions to excite MDs, it is an opposite situation of mapping intersection of Bragg cones by the incident X-rays \cite{sm96,mh97,la98,la99}. In general, only superposition of intensities are to be considered at the intersections of DMS lines, without interference effects, as the intensities on different lines comes from uncorrelated DS sources at different positions of the reciprocal space, as illustrated in Fig.~\ref{fig:ewaldconstruction} by the Q and Q$^\prime$ bright cones intersection.

\begin{figure*}
\includegraphics[width= .75\textwidth]{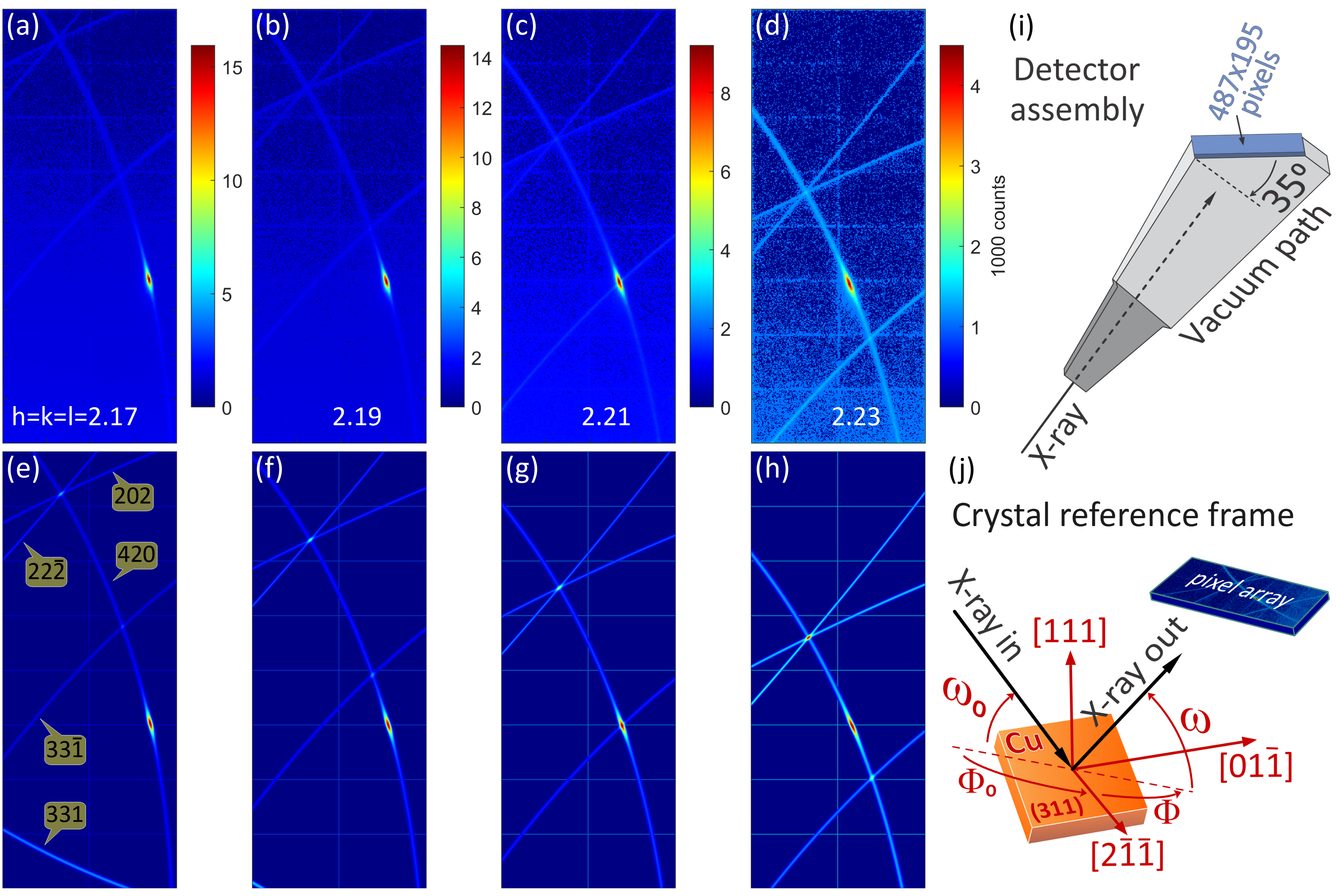}\\
\caption{(a-d) Experimental observation and (e-h) simulation of DMS lines in Cu (311) single crystal with high-flux synchrotron X-rays of 7.82\,keV ($\lambda = 1.585486$\,\AA), $\sigma$-polarization (vertical scattering plane). (i) Detector assembly with pixel array rotated by $35^\circ$ and centered at 565\,mm from the sample; Diamond Light Source beamline I16 \cite{sc10}, pilutus 100k detector. (j) Crystal reference frame. $\omega_0$ and $\omega$ are both equal to the Bragg angle $\theta_{\rm hkl}$ with ${\rm h}={\rm k}={\rm l}$ in the range from 2.17 to 2.23 as indicated in the top panels. $\Phi_0 = -134.56^\circ$ and $\Phi = \Phi_0 + 180^\circ$ in all cases. DMS line indices are shown in the bottom left panel. Simulation is based on isotropic diffuse scattering to light up the DMS lines plus mosaicity at reflection $\bar{2}02$ to account for the behaviour of the intensity spot (red spot) on the 420 DMS line; Complete image data sets are available in SI.}
\label{fig:dmsCuNisbet} 
\end{figure*}

\begin{figure*}
\includegraphics[width=.75\textwidth]{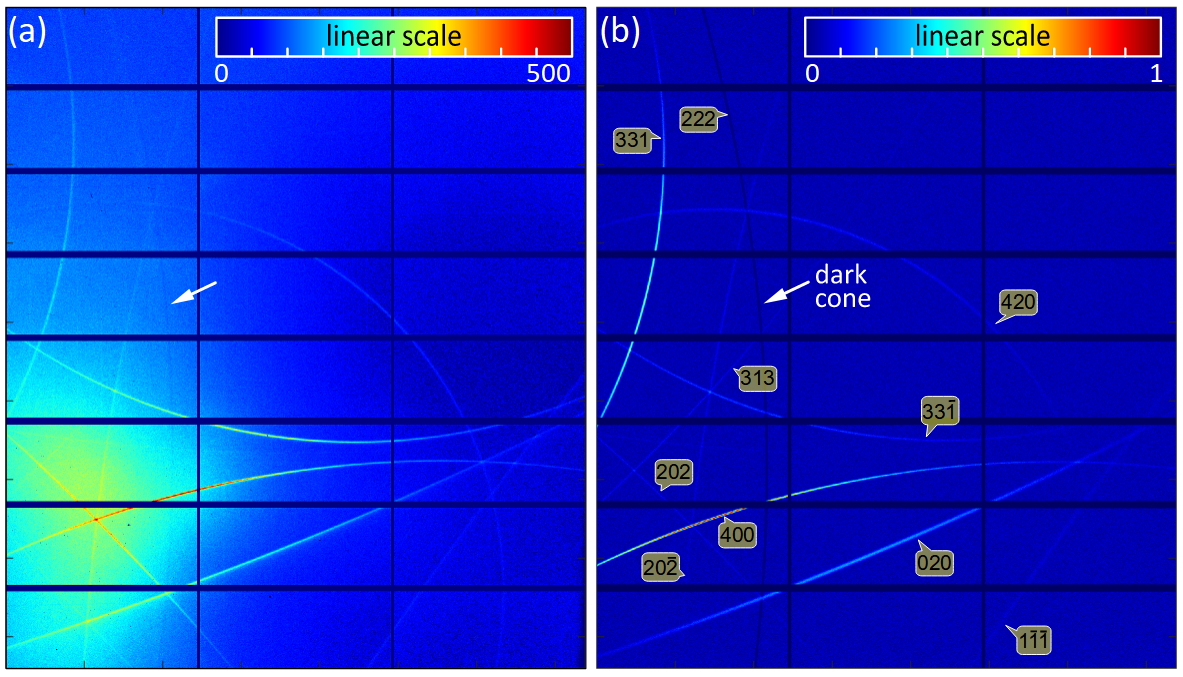}\\
\caption{(a) Experimental observation and (b) simulation of DMS lines in Cu (311) single crystal with high-flux synchrotron X-rays of 7.82\,keV, $\pi$-polarization (horizontal scattering plane). A dark cone is also visible in the imaged area (white arrows). Detector area perpendicular and centered at 850\,mm from the sample; Diamond Light Source I16 \cite{sc10}, pilatus 2M detector. $\omega_0 = \omega = 45^\circ$, corresponding to ${\rm h}={\rm k}={\rm l}=1.8618$, $\Phi_0 = 216^\circ$, and $\Phi = \Phi_0 - 180^\circ$ in the reference frame of Fig.~\ref{fig:dmsCuNisbet}(i). Simulation based on a short-range DS model plus mosaicity, see main text for details.}
\label{fig:2M} 
\end{figure*}

Observation of long DMS lines in perfect crystals is possible due to DS from thermal vibrations \cite{mh99}, as in materials of high thermal conductivity such as copper. Figs.~\ref{fig:dmsCuNisbet}(a)-\ref{fig:dmsCuNisbet}(d) and Fig.~\ref{fig:2M}(a) show experimental DMS lines in Cu (311) single crystal obtained either on vertical or horizontal scattering planes, $\sigma$ or $\pi$ polarization, respectively; see SI for the solid angles recorded on each scattering geometry in comparison with DMS lines on a wide solid angle. Although most of the lines are easily simulated within the isotropic DS model, there are a few intensity features demanding more complex models than accomplished by the DS or CTR models in Eq.~(\ref{eq:W}). In particular, the well defined intensity spot on the 420 DMS line, Figs.~\ref{fig:dmsCuNisbet}(a)-\ref{fig:dmsCuNisbet}(d), that is observed to move along the [111] direction exactly as expected for a CTR in $\theta/2\theta$-scans of symmetrical reflections (specular reflection geometry). However, the direction normal to the sample surface is the [311] direction rather than [111], as depicted in Fig.~\ref{fig:dmsCuNisbet}(j).   

\begin{figure}
\includegraphics[width=0.48\textwidth]{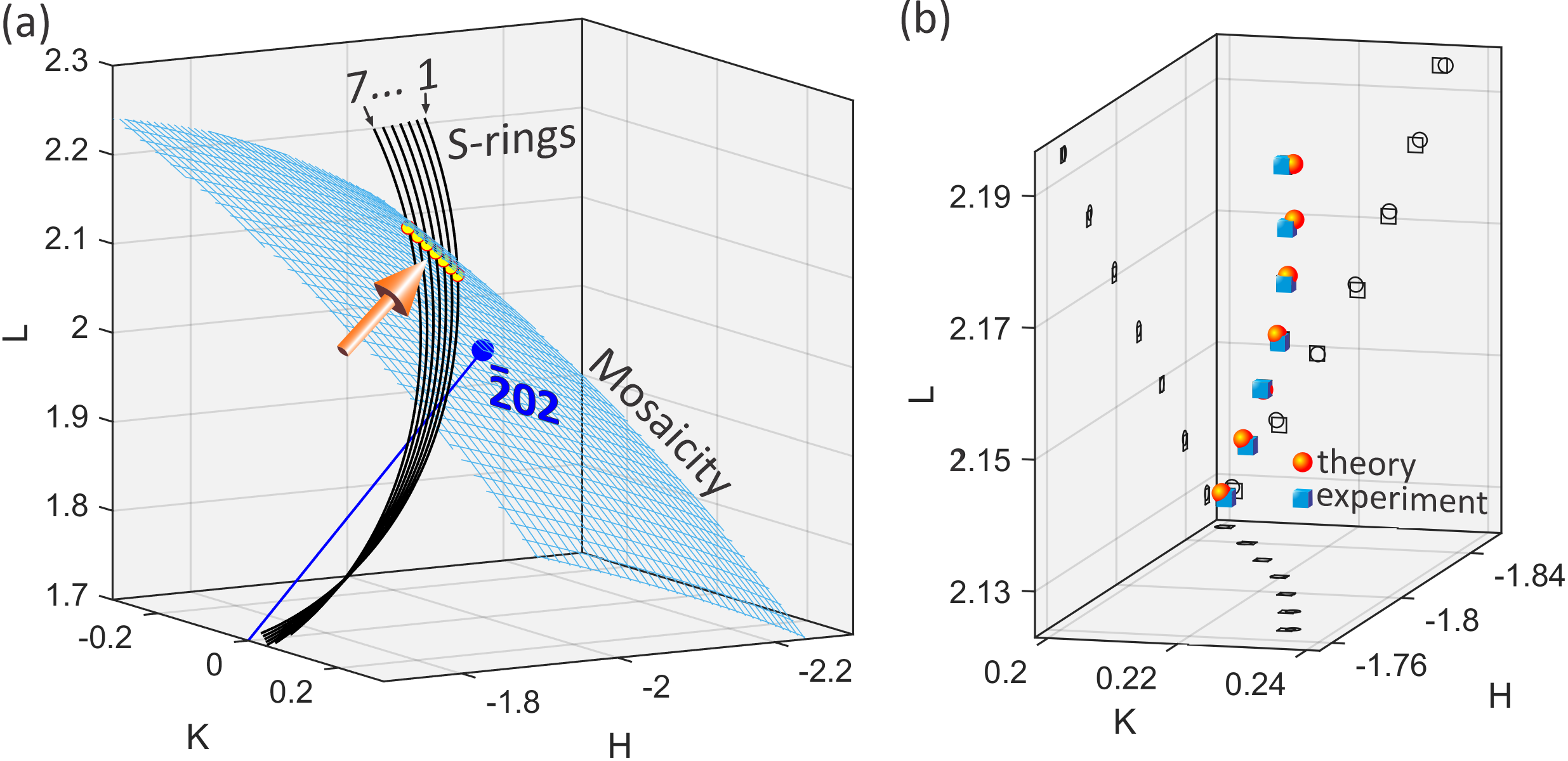}\\
\caption{(a) Three surface intersection spots (big arrow) between the 420 dark cone and Ewald sphere (S-rings) and a spherical shell of radius $\sqrt{8}$ (mesh surface) as standing for mosaicity at around the $\bar{2}02$ node. S-rings 1 to 7 correspond to changes in the incidence angle $\omega_0 = \theta_{\rm hkl}$ with ${\rm h}={\rm k}={\rm l}$ varying from 2.17 to 2.23 in steps of 0.01, $\Phi_0 = -134.56^\circ$, and $\Phi = \Phi_0 + 180^\circ$. (b) Theoretical (red spheres) and experimental (blue cubes) intersection spots of the S-rings with the shell of grain misorientation. Projections of the interception points on the HK, HL and KL planes (open circles and open rectangles in dark gray) are also shown.}
\label{fig:SvectorMosaicity} 
\end{figure}

After transforming each pixel of the detector area, Fig.~\ref{fig:dmsCuNisbet}(e), on HKL coordinates around the 420 S-ring, the scattering vector responsible for the spot intensity ends at coordinates like ${\rm h}^\prime=-1.7988$, ${\rm k}^\prime=0.2246$, and ${\rm l}^\prime=2.1703$ on the S-ring, whose modulus square is very close to 8. The experimental data were then exactly simulated in Figs.~\ref{fig:dmsCuNisbet}(f) - \ref{fig:dmsCuNisbet}(i) by taking the source of spot intensity as located at the intersection between the 420 S-ring and a sphere of radius $\sqrt{8}$, as show in Fig.~\ref{fig:SvectorMosaicity}(a). The perfect agreement between experimental and theoretical sets of HKL coordinates is demonstrated in Fig.~\ref{fig:SvectorMosaicity}(b) where the former set was obtained from the central pixel of the spot in each experimental image, see SI, according to:
\begin{center}
\fbox{pixel ($m,n$)} $\rightarrow$ \fbox{$\omega,\Phi$} $\rightarrow$ \fbox{$\bfm{k}_b$} $\rightarrow$ \fbox{$\bfm{k}_d = \bfm{k}_b - \bfm{Q}$} $\rightarrow$ \fbox{$\bfm{S} = \bfm{k}_d - \bfm{k}$} $\rightarrow$ \fbox{${\rm h}^\prime{\rm k}^\prime{\rm l}^\prime$}\,, 
\end{center}
while the later set was calculated by the intersection of three surfaces: 
\begin{center}
    \fbox{Q dark cone} $\cap$ \fbox{Ewald sphere} ($=$ S-ring) $\cap$ \fbox{sphere of radius $\left|\bfm{H}\right|$}\,.
\end{center}
$\bfm{Q}$ and $\bfm{H}$ stand for diffraction vectors of reflections 420 and $\bar{2}02$, respectively. Shape and relative intensity of the spot were adjusted by adding to $W(\bfm{u})$ in Eq.~(\ref{eq:W}) the term
\begin{equation}\label{eq:mosaic}
    \eta_3 \frac{1}{1 + (\pi N_g \Delta)^2} \exp\left(-\frac{\theta_{S}^2}{2\sigma^2}\right) 
\end{equation}
where $\Delta =(|\bfm{S}|-|\bfm{H}|)/|\bfm{H}|$\; and $\theta_S$ is the angle between $\bfm{S}$ and $\bfm{H}$ vectors, that is, $\Delta = \sqrt{{\rm h}^{\prime 2}+{\rm k}^{\prime 2}+{\rm l}^{\prime 2}}/\sqrt{8} - 1$ and $\cos{\theta_S} = ({\rm l}^\prime-{\rm h}^\prime)\left[2\, ({\rm h}^{\prime 2}+{\rm k}^{\prime 2}+{\rm l}^{\prime 2})\right]^{-1/2}$ as copper is a cubic crystal. The simulations in Figs.~\ref{fig:dmsCuNisbet}(e)-\ref{fig:dmsCuNisbet}(h) were obtained by using $\eta_1=1$, $N=4/\pi$, $\eta_2=0$, $\eta_3=40$, a grain size $N_g = 242$ Bragg planes (31\,nm), and a mosaic spread $\sigma = 3.2^\circ$. As the grain size is the only parameter determining the spot shape along the DMS line, it is the most accurate value with uncertainty about 20\%. The other parameter values are related to each other and, therefore, adjustable within wide ranges capable of generating similar images.

Fig.~\ref{fig:2M}(a) shows the actual lengths of DMS lines tracked on a wide-area detector in horizontal scattering geometry, $\pi$-polarization. By adjusting the scattering angle to $2\theta=\omega_0+\omega \simeq 90^\circ$, X-ray photons from single scattering events are strongly suppressed by polarization near the center of the detection area, enhancing the visibility of DMS lines. It reveals that most of the lines are much shorter than expected within the long-range isotropic DS model in Eq.~(\ref{eq:W}). A short-range model, obtained by  exchanging $1/\left[1 + (\pi N u)^2\right]$ for $\exp\!\left[-(\pi N u)^2 \ln(2)\right]$ in Eq.~(\ref{eq:W}), was used instead to better reproduce the length of the lines. According to the theoretical approach in Eq.~(\ref{eq:DMSlineintensity}), the 400 line is expected to be 40\% weaker that the 331 line and 80\% weaker than the 020 line when considering only the nearest nodes of the respective dark cones as $\left|F_{400}\right|^2\left|F_{\bar{2}22}\right|^2 \simeq 0.6 \left|F_{331}\right|^2\left|F_{\bar{1}\bar{1}1}\right|^2 \simeq 0.2 \left|F_{020}\right|^2\left|F_{202}\right|^2$. Simulation of this extra intensity of the 400 lines was accomplished by adding mosaicity to the $\bar{2}22$ node, as in Eq.~(\ref{eq:mosaic}). For the other nodes, mosaicity contributions appear further way of the imaged area. $\eta_1 = 1$, $N=5/\pi$, $\eta_2=0$, $\eta_3 = 2$, $\theta_S=0$, and $N_g = 11$ were the parameter values used for the simulation shown in Fig.~\ref{fig:2M}(b). A random value between zero and 0.1 was added as statistical noise to all pixels, except for those at the position of the 222 line (white arrow).

\begin{figure}
\includegraphics[width=.48\textwidth]{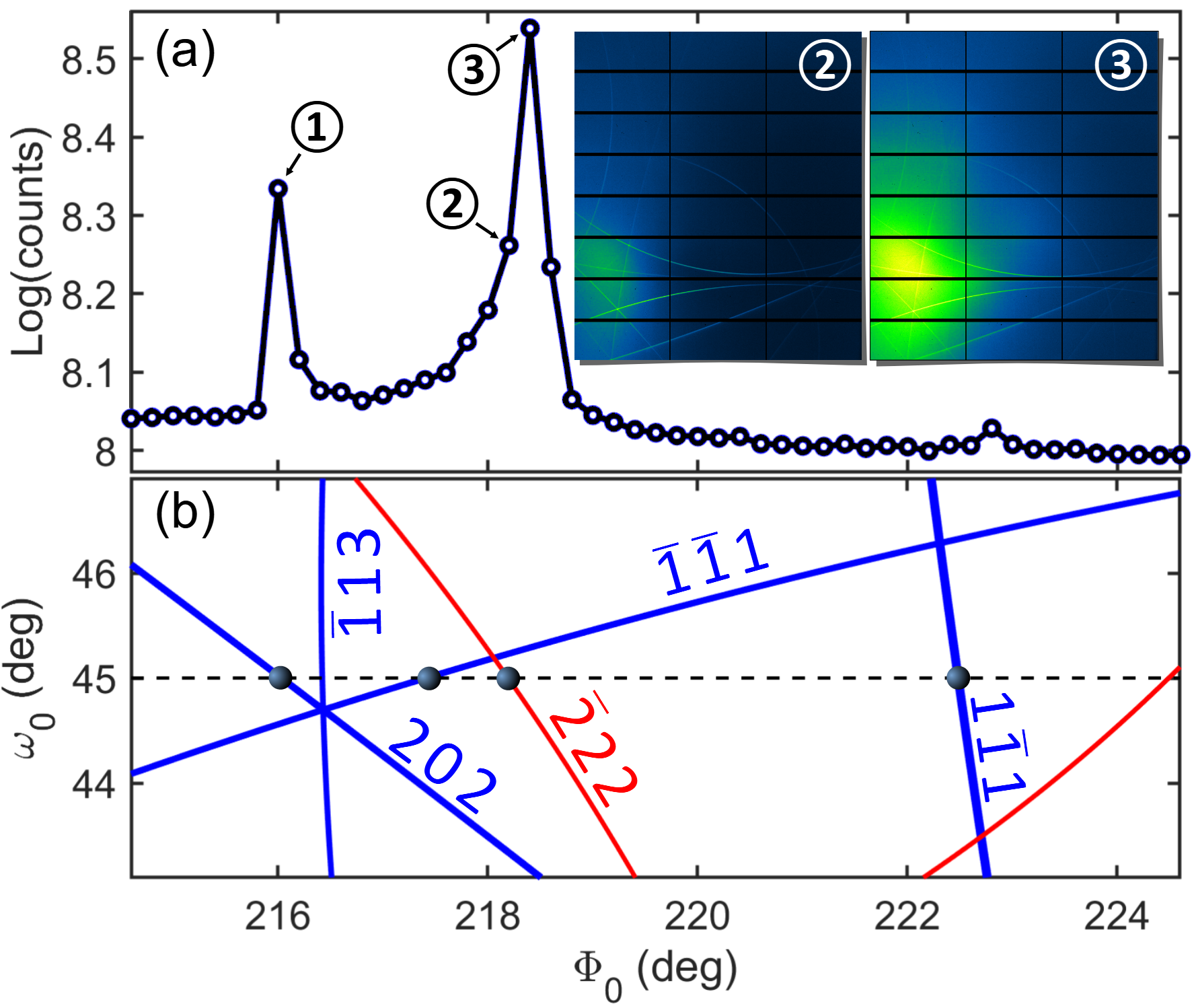}\\
\caption{(a) Integrated detector intensity vs. sample azimuthal angle $\Phi_0$ around [111]. Images at select azimuths (arrows 2 \& 3) are shown as insets and in Fig.~\ref{fig:2M}(a) (arrow 1). (b) 2D representation of Bragg cones vs. incidence angles $\omega_0$ and $\Phi_0$. Intersections with the fixed value of $\omega_0=45^\circ$ (dashed line) indicate azimuths where Bragg-DS channels are excited.}
\label{fig:intint} 
\end{figure}

The 222 bright cone, devoid of intensity contributions from DS sources, allows the $\bar{2}\bar{2}\bar{2}$ dark cone (white arrows) to become visible in areas detecting nebulous scattering from Bragg-DS channels. In Fig.~\ref{fig:intint}, Bragg reflections involved in these channels are identified by plotting the total integrated intensity across the entire detector area as a function of the sample's azimuth, in comparison to a 2D graphic representation of Bragg cones as a function of incidence angles $\omega_0$ and $\Phi_0$. A stronger Bragg-DS channel, distinct from the one in Fig.~\ref{fig:2M}(a) (caused by the 202 Bragg reflection), is excited at a slightly different azimuth, $\Phi_0=218.4^\circ$ (inset of Fig.~\ref{fig:intint}). Observation of these second-order scattering channels provides further evidence of DS and/or mosaicity around reciprocal lattice nodes. Their image corresponds to the intersection of the Ewald sphere associated with the Bragg-diffracted beam with the distribution of intensity in reciprocal space, analogous to the case of the incident X-ray beam's Ewald sphere. However, Bragg-DS channels are highly azimuth-dependent, precluding their use in 3D reciprocal space mapping applications.

\section{Conclusions}
In conclusion, this work has introduced a robust theoretical framework for simulating X-ray diffuse multiple scattering lines, filling a significant gap in existing methodologies and providing a valuable tool for the detailed analysis of experimental data.  The successful application to both silicon and copper single crystals, accounting for isotropic diffuse scattering and crystal truncation, demonstrates the broad applicability of this approach.  Furthermore, the inclusion of grain misorientation effects highlights the potential of this framework to elucidate complex microstructural features.  This work opens a new avenue for investigating diffuse scattering phenomena and their relationship to material properties, with the promise of furthering our understanding of crystal defects and their influence on material behavior.

\begin{acknowledgments}
Acknowledgments are due to the Brazilian funding agencies FAPESP (Grant Nos. proc2023/10775-1, 2022/09531-8,  and 2021/01004-6) and CNPq (Grand No. 310432/2020-0).
\end{acknowledgments}

\end{document}